\documentstyle[twocolumn,aps,epsf]{revtex}
\begin{document}
\draft

\twocolumn[\hsize\textwidth\columnwidth\hsize\csname @twocolumnfalse\endcsname

\title{Prediction of Orbital Ordering in Single-Layered Ruthenates}

\author{Takashi Hotta}
\address{Advanced Science Research Center,
Japan Atomic Energy Research Institute, Tokai, Ibaraki 319-1195,
Japan}

\author{Elbio Dagotto}
\address{National High Magnetic Field Laboratory,
Florida State University, Tallahassee, FL 32306}

\date{\today}

\maketitle

\begin{abstract}

The key role of the orbital degree of freedom to understand
the magnetic properties of layered ruthenates is here discussed.
In the G-type antiferromagnetic phase of Ca$_2$RuO$_4$,
recent X-ray experiments reported the presence of 0.5 hole per site
in the $d_{\rm xy}$ orbital, while the $d_{\rm yz}$ and $d_{\rm zx}$
orbitals contain 1.5 holes.
This unexpected $t_{\rm 2g}$ hole distribution is explained
by a novel state with
orbital ordering (OO), stabilized by a combination of Coulomb
interactions and lattice distortions. 
In addition, the rich phase diagram presented here
suggests the possibility of large magnetoresistance
effects, and predicts a new ferromagnetic OO phase in ruthenates.

\end{abstract}



\vskip2pc]
\narrowtext


The single-layered ruthenate Sr$_2$RuO$_4$ has recently attracted much 
attention, both in its experimental and theoretical aspects, 
since it is an exotic material exhibiting triplet superconductivity
in the solid state \cite{Maeno}.
By analogy with superfluidity in $^3$He, the triplet superconductivity
is believed to originate from ferromagnetic (FM) spin fluctuations 
enhanced by the Hund coupling \cite{Rice}.
Although Knight shift experiments confirm spin triplet pairing
\cite{Ishida}, the symmetry of Cooper pairs remains controversial,
since some experimental results are consistent with
the existence of a line-node gap \cite{gap}, 
in contrast to the theoretical candidate with a
nodeless $p$-wave gap.

In addition to the triplet superconductivity, ruthenates exhibit complex
magnetic properties.
When Sr is partially substituted by Ca, superconductivity is rapidly
destroyed and a paramagnetic (PM) metallic phase appears, while
for Ca$_{1.5}$Sr$_{0.5}$RuO$_4$, a nearly FM metallic phase has been
suggested.
Upon further substitution, the system eventually transforms into
an AFM insulator \cite{Nakatsuji}.
The G-type AFM phase in $\rm Ca_2RuO_4$ is characterized as a standard N\'eel
state with spin $S$=1 \cite{Nakatsuji2,Braden}.
To understand the N\'eel state observed in experiments,
one may consider the effect of the tetragonal
crystal field, leading to the splitting between xy and \{yz,zx\}
orbitals, where the xy-orbital state is lower in energy than the other levels.
When the xy-orbital is fully occupied, 
a simple superexchange interaction at strong Hund coupling
can stabilize the AFM state.
However, recent X-ray absorption spectroscopy studies have shown that
0.5 holes per site exist in the xy-orbital, while 1.5 holes are contained
in the zx- and yz-orbitals \cite{Mizokawa}, suggesting that the above
naive picture based on crystal field effects is incomplete.
This fact suggests that the orbital degree of freedom may play
a more crucial role in the magnetic ordering in ruthenates than
previously anticipated.

In this Letter, the multi-orbital Hubbard model tightly coupled
to lattice distortions is analyzed using numerical and mean-field 
techniques.
Since this model includes several degrees of freedom 
(2-spin and 3-orbital/electron),
it is quite difficult to study large-size clusters.
However, it is believed that the essential character of the competing states
can be captured using a small cluster, 
through the combination of the Lanczos method 
and relaxational techniques.
Mean-field approximations complement and support
the results obtained numerically.
An important conclusion of our analysis
is that the G-type AFM phase is stabilized only
when {\it both} Coulombic and phononic interactions are taken into account.
The existence of a novel orbital ordering is crucial
to reproduce the peculiar hole arrangement observed in 
experiments \cite{Mizokawa}.
Another interesting consequence of the present study is the possibility
of large magneto-resistance phenomena in ruthenates, since 
in our phase diagram the ``metallic'' FM phase is adjacent to the
``insulating'' AFM state.
This two-phase competition is at the heart of 
Colossal Magneto-Resistance (CMR) in manganites \cite{Dagotto}, and, thus,
CMR-like phenomenon could also exist in ruthenates.


Let us briefly discuss a localized Ru$^{4+}$ ion,
in which four electrons are contained in the $4d$ orbitals.
Since the crystal field splitting between $e_{\rm g}$ and $t_{\rm 2g}$
orbitals is larger than the Hund's rule coupling, the Ru$^{4+}$ ion is in
the low-spin state ($S$=1):
three up and one down (or three down and one up) electrons occupy
the triply degenerate $t_{\rm 2g}$ manifold, spanned by 
$d_{\rm xy}$, $d_{\rm yz}$, and $d_{\rm zx}$.
These $t_{\rm 2g}$ electrons can move to the neighboring Ru sites
via the oxygen 2p$\pi$ orbitals, and, in addition, they are correlated 
with each other, and are coupled to the distortion of the RuO$_6$ octahedron.
The above situation is described by the Hamiltonian
$H$=$H_{\rm kin}$+$H_{\rm el-el}$+$H_{\rm el-ph}$.
The first term $H_{\rm kin}$
denotes the hopping of $t_{\rm 2g}$ electrons,
given by
\begin{eqnarray}
  H_{\rm kin} &=& -\sum_{{\bf ia}\gamma \gamma'\sigma}
  t^{\bf a}_{\gamma \gamma'} d_{{\bf i} \gamma \sigma}^{\dag}
  d_{{\bf i+a} \gamma' \sigma},
\end{eqnarray}
where $d_{{\bf i} \gamma \sigma}$ is the annihilation operator
for a $t_{\rm 2g}$-electron with spin $\sigma$ in the $\gamma$-orbital
at site {\bf i} ($\gamma$=xy, yz, and zx),
{\bf a} is the vector connecting nearest-neighbor sites, and
$t^{\bf a}_{\gamma \gamma'}$ is the nearest-neighbor hopping 
amplitude between $\gamma$- and $\gamma'$-orbitals along the 
{\bf a}-direction via the oxygen 2p$\pi$ bond,
given by $t^{\bf x}_{\rm xy,xy}$=$t^{\bf x}_{\rm zx,zx}$
=$t^{\bf y}_{\rm xy,xy}$=$t^{\bf y}_{\rm yz,yz}$=$t$=1,
and zero otherwise.
Note that only the two-dimensional case is considered throughout
this paper.

The second term $H_{\rm el-el}$ denotes the Coulomb interactions between
$t_{\rm 2g}$ electrons, given by
\begin{eqnarray}
  H_{\rm el-el} &=& U \sum_{{\bf i},\gamma}
  \rho_{{\bf i}\gamma\uparrow} \rho_{{\bf i}\gamma\downarrow}
  + U'/2 \sum_{{\bf i}\gamma \ne \gamma'} 
  \rho_{{\bf i}\gamma} \rho_{{\bf i}\gamma'}
  \nonumber \\
  &+& J/2 \sum_{{\bf i},\sigma,\sigma',\gamma \ne \gamma'}
  d_{{\bf i}\gamma\sigma}^{\dag}
  d_{{\bf i}\gamma'\sigma'}^{\dag}
  d_{{\bf i}\gamma\sigma'}
  d_{{\bf i}\gamma'\sigma}
  \nonumber \\
  &+& J'/2 \sum_{{\bf i},\sigma \ne \sigma',\gamma \ne \gamma'}
  d_{{\bf i}\gamma\sigma}^{\dag}
  d_{{\bf i}\gamma\sigma'}^{\dag}
  d_{{\bf i}\gamma'\sigma'}
  d_{{\bf i}\gamma'\sigma},
\end{eqnarray}
where
$\rho_{{\bf i}\gamma\sigma}$=
$d_{{\bf i}\gamma\sigma}^{\dag} d_{{\bf i}\gamma\sigma}$,
$\rho_{{\bf i}\gamma}$=
$\sum_{\sigma}\rho_{{\bf i}\gamma\sigma}$,
$U$ is the intra-orbital Coulomb interaction,
$U'$ is the inter-orbital Coulomb interaction,
$J$ is the inter-orbital exchange interaction,
and $J'$ is the pair-hopping amplitude between different orbitals. 
Note that $U$=$U'$+$J$+$J'$ due to rotational invariance in orbital space,
and the relation $J'$=$J$ holds from the evaluation of Coulomb
integrals \cite{Dagotto}.

Finally, the third term $H_{\rm el-ph}$ indicates the electron-lattice
coupling, expressed as \cite{Struge}
\begin{eqnarray}
  H_{\rm el-ph} &=& g \sum_{\bf i}
   (Q_{z{\bf i}} \rho_{{\bf i}{\rm xy}}
   +Q_{x{\bf i}} \rho_{{\bf i}{\rm yz}}
   +Q_{y{\bf i}} \rho_{{\bf i}{\rm zx}}) \nonumber \\
   &+& (k/2) \sum_{\bf i} (Q_{2{\bf i}}^2 + Q_{3{\bf i}}^2),
\end{eqnarray}
where $g$ is the electron-lattice coupling constant,
$k$ is the spring constant,
$Q_{x{\bf i}}$=$(-1/2)Q_{3{\bf i}}$+$(\sqrt{3}/2)Q_{2{\bf i}}$,
$Q_{y{\bf i}}$=$(-1/2)Q_{3{\bf i}}$$-$$(\sqrt{3}/2)Q_{2{\bf i}}$,
and $Q_{z{\bf i}}$=$Q_{3{\bf i}}$.
Here $Q_{2{\bf i}}$ and $Q_{3{\bf i}}$ are the $(x^2-y^2)$- and
$(3z^2-r^2)$-mode distortions of the RuO$_6$ octahedron, respectively.
The self-trapping energy is defined as $E_{\rm ph}$=$g^2/(2k)$.
Since oxygens are shared by adjacent octahedra, the distortions
do not occur independently, and 
a cooperative treatment should be employed in this problem.
A simple way to include such an effect is to optimize directly
the displacement of oxygen ions \cite{Dagotto}.
In practice, considering sites ${\bf i}$ and ${\bf i+a}$, the oxygen
in between is only allowed to move {\it along} the ${\bf a}$-axis
(neglecting buckling and rotations), while
apical oxygens move freely along the $z$-direction.


The above model is believed to provide a good starting point to study
the electronic properties of ruthenates, but it is difficult
to solve even approximately.
To gain insight into this complex system, an unbiased technique
should be employed first. Thus, in this paper, first a small 2$\times$2 
plaquette cluster is analyzed in detail
by using the Lanczos algorithm for the exact diagonalization, and
the relaxation technique to determine the oxygen positions.
In actual calculations, at each step for the relaxation, 
the electronic portion of the Hamiltonian is exactly
diagonalized for a fixed distortion.
Iterations are repeated until the system converges to the global ground
state \cite{note:ed}.

\begin{figure}[t]
\centerline{\epsfxsize=8.truecm \epsfbox{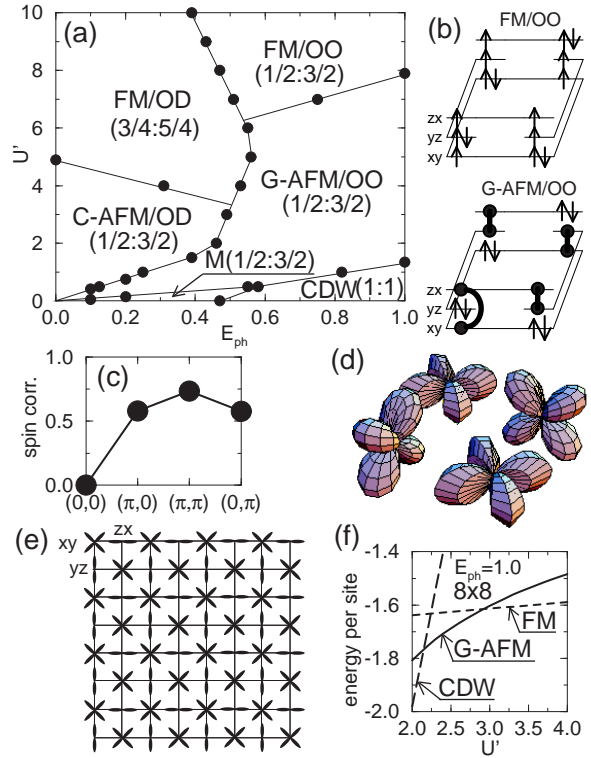} }
\label{fig1}
\smallskip
\caption{
(a) Ground state phase diagram for the full model with $J$=$3U'/4$.
The notation is explained in the text. Other choices of $J$ lead to a
similar phase diagram (see also Fig.2).
(b) Electron configurations in $t_{\rm 2g}$ orbitals for the FM/OO
and G-AFM/OO states. Solid circles connected by solid curves denote 
the local $S$=1 spin. Thin lines connecting orbitals denote the
allowed hopping processes.
(c) Spin correlation for the G-AFM phase.
(d) Orbital pattern for FM/OO and G-AFM/OO states.
(e) Orbital ordering in ruthenates, deduced from the 2$\times$2 result.
(f) Mean-field energies for CDW, G-AFM, and FM states on
the 8$\times$8 lattice and $E_{\rm ph}$=1.0,
with lattice distortions assumed from the exact 2$\times$2 results.
}
\end{figure}


The main result of this paper is summarized in the ground state phase diagram
shown in Fig.~1(a).
There are six phases in total, which are divided into two groups.
One group is composed of phases stemming from
the $U'$=0 or $E_{\rm ph}$=0 limits.
The origin of these phases will be addressed later, but first
their main characteristics are briefly discussed.
For $E_{\rm ph}$=0, a C-type AFM orbital disordered (OD)
phase appears in the region of small and intermediate $U'$. 
This state is characterized by
$n_{\rm xy}$:$n_{\rm yz}$+$n_{\rm zx}$=1/2:3/2,
where $n_{\gamma}$ is the hole number per site at the $\gamma$-orbital.
Hereafter, a shorthand notation such as ``1/2:3/2'' is used to denote the
hole configuration. For large $U'$, and still $E_{\rm ph}$=0, a FM/OD phase
characterized by 3/4:5/4 is stable, which may correspond to Sr$_2$RuO$_4$. 
On the other hand, for $U'$=0 and small $E_{\rm ph}$, a ``metallic'' (M)
phase with small lattice distortion is observed,
while for large $E_{\rm ph}$,
a charge-density-wave (CDW) state characterized by 1:1 was found.
In short, the G-type AFM phase observed experimentally \cite{Braden}
does {\it not} appear, neither for $E_{\rm ph}$=0 nor for $U'$=0.

Another group includes two phases which are not connected to either
$E_{\rm ph}$=0 or $U'$=0.
It is only in this group, with {\it both} lattice and Coulomb effects being
relevant, that for intermediate $U'$ the G-type AFM and orbital
ordered (OO) phase with 1/2:3/2 found in experiments \cite{Mizokawa} 
is stabilized. At larger $U'$, a FM/OO phase 
occurs with the same hole arrangement.
The electron configurations for the FM and AFM phases are summarized
in Fig.~1(b).
In the FM phase, since an $S$=1 spin with $S_z$=+1 is formed at each site,
the up-spin number is unity at each orbital,
while the down-spin distribution depends on the orbital.
In the AFM state, the configuration of double-occupied
orbitals is the same as in the FM phase,
but the single-occupied orbital contains 0.5 up- and 0.5-down spins
on average,
since the $S$=1 spin direction fluctuates due to the AFM
coupling between neighboring $S$=1 spins.
However, as shown in Fig.~1(c), the spin correlations peak at
($\pi$,$\pi$), indicating the G-AFM structure.
Except for the spin direction, the charge and orbital configuration
in the FM/OO phase is the same as in the G-AFM/OO state.
As shown in Fig.~1(d), a clear 
ordering pattern including xy, yz, and zx orbitals
is suggested for these FM and AFM phases \cite{note:OO}.
Deduced from the present 2$\times$2 result, the proposed bulk
orbital ordering in ruthenates is shown in Fig.~1(e).
Note that in this pattern xy-orbitals occupy only half the Ru-sites,
while zx- (yz-) orbitals connect second nearest
neighbor xy-orbitals along the x- (y-) axis.
This structure, quite different from a uniform population of the
xy-orbitals, is natural from the viewpoint of the kinetic
energy gain of $t_{\rm 2g}$ electrons, with all orbitals contributing
to such gain.


Although the results discussed above are exact, they have been obtained
using a small-size cluster. Unfortunately, larger clusters cannot be
studied due to severe limitations of memory and CPU time.
However, it is possible to consider larger lattices by employing
a mean-field approximation.
In the actual calculations, only the diagonal portion of the Coulomb
interactions in spin and orbital space are included by carrying out
a standard Hartree-Fock decoupling procedure,
working with the fixed lattice distortions deduced from the 
2$\times$2 results.
In Fig.~1(f), the mean-field energies vs. $U'$ on the 8$\times$8 cluster
are depicted at $E_{\rm ph}$=1.0.
It is remarkable that with increasing $U'$, the ground state changes
from CDW, to G-AFM, and finally to FM, in excellent 
agreement with the 2$\times$2 results \cite{note:MFA}.
For intermediate values of $U'$, the G-type AFM phase with the hole
arrangement observed in experiments is easily obtained,
if the lattice distortion to produce
the orbital ordering in Fig.~1(e) is assumed.
For large $U'$, on the other hand, the FM phase is stabilized (this
phase has not been observed experimentally yet, and it may correspond to
a FM insulator).
The mean-field generates stable phases and their competition are
in qualitative agreement with 
the small-cluster results for $E_{\rm ph}$$\ne$0 and $U'$$\ne$0,
suggesting that the orbital ordering coupled to lattice distortions
is essential to explain the magnetic structure of ruthenates.


Now let us discuss intuitively the origin of the
complex OO pattern observed here.
Due to the Hund coupling, a local $S$=1 spin is formed
by the four electrons, as shown in Fig.~1(b).
Note that each orbital includes at least one spin-majority electron,
leading to the cancellation of their distortions due to the relation
$Q_{x{\bf i}}$+$Q_{y{\bf i}}$+$Q_{z{\bf i}}$=0.
Thus, the OO pattern is determined by the choice of orbital occupancy
of the excess spin-minority electron.
Note that the energy levels of orbitals occupied and unoccupied 
by this excess electron are
$-2E_{\rm ph}$ and $+E_{\rm ph}$, respectively.
The two-dimensional network composed of xy-orbitals is occupied
in a {\it bipartite} manner by the excess electrons, to gain kinetic energy
(namely, to avoid the total ``freezing'' of the xy-orbitals).
Other excess electrons occupy yz- and zx-orbitals, 
with mobility only along one-dimensional (1D) chains,
coupled to $Q_{x{\bf i}}$ and $Q_{y{\bf i}}$ distortions, respectively.
These distortions should occur in pairs to cancel the
in-plane total distortions, since $Q_{\rm x}$ ($Q_{\rm y}$) denotes
the elongation along the x- (y-) direction.
Then, half of 1D chains of yz- and zx-orbitals are occupied, to
avoid the sites with occupied xy-orbitals.
The above considerations naturally lead to the orbital pattern
Fig.~1(e).


Another interesting aspect of the phase diagram Fig.~1(a) is
the appearance of a FM/OD phase adjacent to the G-AFM/OO phase,
suggesting a competition between ``metallic'' FM 
and ``insulating'' AFM states.
Since this two-phase competition is a key concept to understand CMR
manganites \cite{Dagotto},
the present result suggests the possibility of 
CMR-like effects in ruthenates as well.
If a small-radius alkaline-earth ion is substituted
for Ca or Sr, CMR-like phenomena may be observed in Ru-oxides.
Note that a two-phase competition has already been observed
experimentally in the bilayer ruthenates \cite{Cao}.


To complete the intuitive analysis, the cases of $E_{\rm ph}$=0 and
$U'$=$J$=0 are discussed separately.
First, let us consider the pure Coulombic model
$H_{\rm el}$=$H_{\rm kin}$+$H_{\rm el-el}$
using the Lanczos method.
In Fig.~2(a), the ground state phase diagram on the
($U'$$-$$J$, $J$) plane is depicted
by comparing the energies for the states classified
by the $z$-component of total spin.
As discussed in the literature of the multi-band Hubbard model
\cite{Imada}, the FM phase appears for large $J$ and small $U'$$-$$J$.
Although it is difficult to analyze transport properties
of the FM phase in this small cluster calculation,
the FM state characterized by 3/4:5/4 is likely metallic,
since the orbitals are disordered.
In fact, there is no clear peak in 
the orbital correlation as shown in Fig.~2(b).
For large $U'$$-$$J$, the singlet ground state appears,
but this is not the N\'eel, but the C-type AFM state, 
since the spin correlation function $S({\bf q})$ has two peaks
at ${\bf q}$=$(\pi,0)$ and $(0, \pi)$, not at $(\pi,\pi)$,
as shown in the inset of Fig.~2(a).
Note here that this singlet ground state is doubly degenerate,
since there occurs two types of C-AFM states characterized
by the peak of $S({\bf q})$ at ${\bf q}$=$(0,\pi)$ and $(0,\pi)$.
This degeneracy originates from the geometry of the two kinds of
possible 1D bands composed of yz- and zx-orbitals.
Considering one of the FM bonds among the $S$=1 spins of the C-AFM state, 
the excess electron in the yz- (zx-) orbital can
move along the y- (x-) direction, providing a kinetic energy gain,
while a superexchange interaction occurs in the AFM bonds.
Thus, the coexistence of FM and AFM bonds in the C-AFM phase is 
due to the balance of kinetic energy and superexchange interaction.
Since the orbital correlation function in Fig.~2(c) is composed
of several orbital patterns of degenerate states, the C-AFM
phase is also OD.
In short, in the 2$\times$2-plaquette calculation,
the singlet ground state of a pure $H_{\rm el}$ is C-AFM and OD if
lattice distortions are not included, in
disagreement with experiments \cite{Mizokawa}. The G-AFM state is
not stable in a purely Coulombic model due to a Hund coupling energy
penalization for its mobile electrons, compared with the C-AFM state
that has better electronic mobility. 

\begin{figure}[t]
\centerline{\epsfxsize=8.truecm \epsfbox{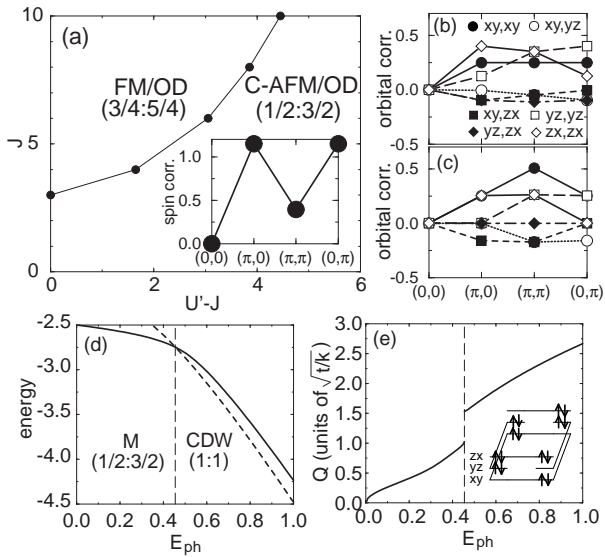} }
\label{fig2}
\smallskip
\caption{(a) Ground state phase diagram for $H_{\rm el}$.
Inset denotes the spin correlation in the C-AFM phase.
(b)-(c) Fourier transform of orbital correlation function
for FM and C-AFM phases, respectively.
(d) Ground state energy vs. $E_{\rm ph}$ for $H_{\rm ph}$.
(e) Average lattice distortion $Q$ vs. $E_{\rm ph}$.
In the inset, the electron configuration for CDW state is shown.
}
\end{figure}


Consider now the pure phononic model
$H_{\rm ph}=H_{\rm kin}+H_{\rm el-ph}$
analyzed by using the relaxation technique.
In Figs.~2(d) and (e), the ground state energy and the average distortion
$Q$ are shown as a function of $E_{\rm ph}$, where $Q$ is defined as 
$Q$=(1/4)$\sum_{\bf i}\sqrt{Q_{2{\bf i}}^2+Q_{3{\bf i}}^2}$.
Note that the unit of $Q$ is $\sqrt{t/k}$, and $k$ is assumed to
be constant here.
For small $E_{\rm ph}$, the ground state with small lattice distortion
is characterized by 1/2:3/2.
When $E_{\rm ph}$ is further increased, the distortion abruptly becomes
larger, and the hole configuration changes to 1:1.
In this state, to gain lattice energy, two of the three $t_{\rm 2g}$
orbitals are doubly occupied, leading to the orbital-dependent CDW state.
The electron configuration is shown in the inset of Fig.~2(e),
indicating that the orbital population changes from site to site.
Note that the ground state of $H_{\rm ph}$ is always non-magnetic
without the electron correlation that forbids double occupancy. Then,
the existence of a finite Hund coupling, and concomitant Hubbard
repulsion, is crucial to reproduce experimental results, as long as it
is supplemented by a lattice distortion.


Summarizing, to understand the magnetic properties of ruthenates,
a model with Coulombic and phononic interactions has been analyzed
by numerical techniques and mean-field approximations.
It has been found that the stabilization of the nontrivial orbital 
ordering state described here is crucial to understand the 
``1/2:3/2'' G-type AFM phase observed in experiments. Both phononic
and Coulombic interactions play an important role in this stabilization. 
In addition, (i) a competition between FM/OD and AFM/OO
phases suggests the possibility of large magnetoresistance effects 
in ruthenates, and (ii) a FM/OO phase is predicted to exist for
ruthenates with large Coulomb interactions.


The authors thank K. Ueda, T. Takimoto, T. Maehira, G. Cao,
and S. Nakatsuji for discussions.
The work has been supported in part
by grant NFS-DMR-9814350 and the In-House Research Program at the NHMFL.

\vskip-0.4cm


\begin{references}

\bibitem{Maeno} Y. Maeno {\it et al}., Nature {\bf 372}, 532 (1994);
Y. Maeno, T. M. Rice, and M. Sigrist, Physics Today {\bf 54}, 42 (2001).

\bibitem{Rice} T. M. Rice and M. Sigrist,
J. Phys.: Condens. Matter. {\bf 7}, L643 (1995).

\bibitem{Ishida} K. Ishida {\it et al.},
Nature {\bf 396}, 658 (1998)

\bibitem{gap} K. Ishida {\it et al.},
Phys. Rev. Lett. {\bf 84}, 5387 (2000).

\bibitem{Nakatsuji} S. Nakatsuji and Y. Maeno,
Phys. Rev. Lett. {\bf 84}, 2666 (2000);
G. Cao {\it et al.}, Phys. Rev. B{\bf 56}, R2916 (1997).

\bibitem{Nakatsuji2} S. Nakatsuji, {\it et al.},
J. Phys. Soc. Jpn. {\bf 66}, 1868 (1997).

\bibitem{Braden} M. Braden {\it et al.},
Phys. Rev. B{\bf 58}, 847 (1998).

\bibitem{Mizokawa} T. Mizokawa {\it et al.},
Phys. Rev. Lett. {\bf 87}, 077202 (2001).

\bibitem{Dagotto} E. Dagotto, T. Hotta, and A. Moreo,
Phys. Rep. {\bf 344}, 1 (2001).

\bibitem{Struge} M. D. Struge,
Solid State Physics {\bf 20}, 91 (1967).

\bibitem{note:ed}
The Hilbert space considered here has 2.5$\times$10$^6$ states, and
a typical relaxation procedure per coupling set involves about 100
iterations, highlighting the magnitude of the computational effort.

\bibitem{note:OO}
A ferro ``0:2'' xy-orbital ordered state was
suggested in V. I. Anisimov {\it et al.}, cond-mat/0011460,
but it is different from experiments in Ref.~\cite{Mizokawa}.
In our calculations this state is an excited state for
$E_{\rm ph}$$\alt$2.0.
Note also that Mizokawa {\it et al.} (Ref.\cite{Mizokawa}) 
suggested a combination of ``complex'' orbitals,
yz and (xy+$i$zx)/$\sqrt{2}$, while the present OO is
a combination of real orbitals.

\bibitem{note:MFA}
The open boundary condition is imposed on the 2$\times$2 cluster, 
while the periodic boundary condition (PBC) is used for the 8$\times$8.
If the PBC is used for the 2$\times$2 plaquette, the energy unit
is multiplied by two, i.e. the values of interactions are reduced by half.
The quantitative difference between 2$\times$2 numerical and 8$\times$8
mean-field results is accounted for by this factor two.

\bibitem{Cao} G. Cao {\it et al.},
Phys. Rev. B{\bf 56}, 5387 (1997).

\bibitem{Imada} M. Imada {\it et al.},
Rev. Mod. Phys. {\bf 70}, 1039 (1998).

\end{references}
\end{document}